\documentclass[11pt]{article}
\usepackage{amsmath}
\usepackage{hyperref}
\usepackage{bm}
\usepackage[numbers,sort&compress]{natbib}
\pdfoutput=1
\begin{document}
\title{Inertial Rise in Short Capillaries}
\author{Orest Shardt, Prashant R. Waghmare, J.J. Derksen, Sushanta K. Mitra \\
\\\vspace{6pt} University of Alberta, Edmonton, AB, T6G 2R3, Canada}
\maketitle
%% The abstract (in this file, and that submitted as text to arXiv) should include the exact phrase
%% "fluid dynamics video" or "fluid dynamics videos"
\begin{abstract}
In this fluid dynamics video we show capillary rise experiments with diethyl ether in short tubes. The height of each short tube is less than the maximum height the liquid can achieve, and therefore the liquid reaches the top of the tube while still rising. Over a narrow range of heights, the ether bulges out from the top of the tube and spreads onto the external wall.
\end{abstract}
% main text
\section*{Description}
% The {\em hyperref} package is used to make links to the videos.
%% The format is: \href{URL of video}{name that will appear in the text}
% Two sample videos are
% \href{http://ecommons.library.cornell.edu/bitstream/1813/8237/2/LIFTED_H2_EMS T_FUEL.mpg}{Video1} and
% \href{http://ecommons.library.cornell.edu/bitstream/1813/8237/4/LIFTED_H2_IEM_FUEL.mpg}{Video2}.

The rise of diethyl ether (density $710\text{ kg/m}^3$, dynamic viscosity $0.22\text{ mPa}\cdot\text{s}$, and surface tension $16.6\text{ mN/m}$) under normal gravity in 0.8 to 1.0~mm diameter cylindrical capillary tubes is oscillatory due to its low viscosity and therefore delayed development of the viscous boundary layer\citep{Quere97}. The liquid achieves a maximum height of 13.5~mm in a 0.8~mm diameter tube that is 16~mm high. When the same experiment is performed in a tube that is only 4.7~mm high, the rising liquid reaches the top of the tube, the meniscus inverts, and a bulge of liquid forms and wets the outer surface. As ether spreads down the outer surface, the ether inside the capillary tube retreats downward and then oscillates vertically. After inverting several times, the meniscus comes to a rest, pinned at the top of the tube.

External wetting occurs when the kinetic energy of the liquid exceeds a reference surface energy. In this case, a reasonable choice is the surface energy of a spherical drop that has the same radius as the capillary tube. If the kinetic energy is too low, the liquid bulges outward, does not spread, and returns into the tube and oscillates. The kinetic energy is insufficient when the height of the liquid column or the rise speed is too low. We show that the ether does not spread onto the outer surface when the tube is 2.7~mm high (too short) or 11.5~mm high (too slow).

Wetting of the external surface also happens with capillary tubes that have a square cross-section (0.8~mm inner side width). We highlight the different stages of the capillary rise phenomenon starting with the surface waves that are produced when the tube touches the ether and ending with the oscillations of the meniscus.

\subsection*{Acknowledgments}
Support from NSERC for O.S. is gratefully acknowledged. The authors thank the Canadian Centre for Welding and Joining (CCWJ) for the use of a high-speed camera.

\end{document}